\documentclass[sn-aps,Numbered,iicol]{sn-jnl}% American Physical Society
\usepackage{graphicx}%
\usepackage{multirow}%
\usepackage{amsmath,amssymb,amsfonts}%
\usepackage{amsthm}%
\usepackage{mathrsfs}%
\usepackage[title]{appendix}%
\usepackage{xcolor}%
\usepackage{textcomp}%
\usepackage{manyfoot}%
\usepackage{booktabs}%
\usepackage{algorithm}%
\usepackage{algorithmicx}%
\usepackage{algpseudocode}%
\usepackage{listings}%
\usepackage{subcaption}
\theoremstyle{thmstyleone}%
\theoremstyle{thmstyletwo}%

\theoremstyle{thmstylethree}%

\begin{document}

\title{On stability issues of the HEOM method}


\author[1]{\fnm{Malte} \sur{Krug}}%\email{malte.krug@uni-ulm.de}

\author*[1]{\fnm{J\"{u}rgen} \sur{Stockburger}}\email{juergen.stockburger@uni-ulm.de}

\affil[1]{\orgdiv{Institute for Complex Quantum Systems and IQST}, \orgname{Ulm University}, \orgaddress{\street{Albert-Einstein-Allee 11}, \postcode{89081} \city{Ulm}, \country{Germany}}}


\abstract{The Hierarchical Equations of Motion (HEOM) method has become one of the cornerstones in the simulation of open quantum systems and their dynamics. It is commonly referred to as a non-perturbative method. Yet there are certain instances where the necessary truncation of the hierarchy of auxiliary density operators seems to introduce errors which are not fully controllable. We investigate the nature and causes of this type of critical error both in the case of pure decoherence, where exact results are available for comparison, and in the spin-boson system, a full system-reservoir model. We find that truncating the hierarchy to any finite size can be problematic for strong coupling to a dissipative reservoir, in particular when combined with an appreciable reservoir memory time.}

\maketitle

\section{Introduction}\label{sec:intro}
The dynamics of open quantum systems is a field in which has seen the development of mature methodologies and powerful abstractions such as completely positive maps~\cite{alick87}, Lindblad generators~\cite{breue02}, path integral methods~\cite{weiss08b} and stochastic approaches~\cite{diosi98,strun99b,stock01,stock02,stock16a}. All these approaches vary in their general applicability, in particular, to non-Markovian dynamics, as well as their potential as computationally feasible methods. Another computational method which has attracted increasing attention in recent years is the hierarchical equations of motion method (HEOM)~\cite{tanim89,tanim91,tanim06,tanim20,xu22}, which extends the dynamics of the reduced density matrix by adding a (potentially large) number of auxiliary density operators (ADOs). Through these, one obtains a time-local version of the open-system dynamics without the usual construction of a dissipation superoperator. The latter approach typically requires application of Born, Markov and secular approximations, yielding a superoperator depending on \emph{both} system and reservoir properties. The HEOM approach, on the other hand, offers a structure where the terms describing dissipation only depend on the system-reservoir coupling. This means that Hamiltonians with non-trivial level structure or complicated explicit time dependence can be adapted in the HEOM formalism far more easily than in master equations. This feature was leveraged in the application of HEOM to multi-dimensional spectroscopy~\cite{ishiz06}. Recent developments greatly extending its utility in the regime of ultralow temperature~\cite{xu22} raise the expectation that this advantage will also be leveraged in the simulation of quantum devices.

All HEOM approaches rest on a multi-exponential decomposition
\begin{equation}\label{eq:C}
    C(t) = \sum_{j=1}^K d_j e^{-z_j t}, \quad t\geq 0
\end{equation}
of the two-time correlation function of the quantum reservoir coupled to the system of interest. In the case of a Gaussian reservoir, it provides all information needed to determine the reduced dynamics. In a strict analytic sense, all thermal (Matsubara) frequencies should be among the $z_j$ of such a decomposition. This suggests that the required number $K$ of exponential terms will rise when lower temperatures are considered, eventually rendering the method infeasible in terms of computational cost. However, when requiring the sum on the right-hand side of eq. (\ref{eq:C}) only to be a close numerical approximation, the number of terms can be drastically lowered~\cite{xu22}, allowing HEOM computations at arbitrary temperature. In the resulting Free-Pole HEOM (FP-HEOM) approach, the coupled dynamics of the reduced density matrix $\rho_{\bf 00}(t)$ and ADOs $\rho_{\bf m,n}(t)$ reads
\begin{align}\label{Eq:baryheom4}
%\begin{split}
\frac{d}{dt} \hat{\rho}_{\bf m,n} &=
-\left(i\mathcal{L}_S
+\sum_{k=1}^{K} m_{k} z_{k} + \sum_{k=1}^{K}
n_{k} z_{k}^{*} \right) \hat{\rho}_{{\bf m,n}} \nonumber\\
&-i\sum_{k=1}^{K}\sqrt{(m_k + 1)d_k} \left[\hat{q},\hat{\rho}_{{\bf m}_k^{+},{\bf n}} \right]  \nonumber\\
&-i\sum_{k=1}^{K} \sqrt{(n_k + 1) d_k^*}\left[\hat{q},\hat{\rho}_{{\bf m,n}_k^{+}} \right]  \nonumber\\
&-i\sum_{k=1}^{K}\sqrt{m_k d_{k}}\;\hat{q}\;\hat{\rho}_{{\bf m}_k^{-},{\bf n}}
\nonumber\\
&+i\sum_{k=1}^{K}
\sqrt{n_k d_{k}^{*}}\;\hat{\rho}_{{\bf m,n}_k^{-}}\hat{q}.
%\end{split}
\end{align}
Here boldface indices $\bf m,n$ are multi-indices $(m_1,\ldots m_K)$, and $(n_1,\ldots n_K)$, and the superscripts ``+`` and ``-'' in conjunction with a subscript $k$ mean that the $k$-th element of $\bf m$ or $\bf n$ is raised/lowered by 1. In FP-HEOM, both the coefficients $d_j$ and the decay rates $z_j$ can be complex numbers, however, the real part of each $z_j$ must be positive for a decaying $C(t)$.

Theoretically, an infinite number of ADOs is required for any HEOM system to be exact, which is clearly impractical. Computationally, equations like (\ref{Eq:baryheom4}) are truncated to systems of equations with a finite number of ADOs. Typically, it is tacitly assumed that increasing the number of ADOs will allow the result to converge. This seems to be the case in the majority of applications. However, cases are reported~\cite{shi18,dunn19,yan20,mengpriv}, where convergence seems to be very slow, and only pointwise with respect to time. In a few examples, the limit of large truncation level even appears to be divergent. The aim of the following sections is to develop a better understanding of the convergence of HEOM dynamics with respect to truncation depth and to point out potential remedies to convergence problems.

\section{Pure dephasing in the hierarchy formalism}
\subsection{Analytic results}

For pure dephasing, an analytic result in terms of quadratures of the reservoir correlation function is available,
\begin{equation}
    \langle q |\hat{\rho}(t)| q' \rangle =  
    \exp(- (q-q')^2\phi(t)) \langle q |\hat{\rho}(0)| q' \rangle
\end{equation}
with real function
\begin{equation}
    \phi(t) = \int\limits_0^t ds \int\limits_0^s ds' C(s-s')
    = \int\limits_0^t d\tau (t-\tau) C(\tau).
\end{equation}
In the Markovian limit, $\phi(t)$ rises linearly with time, the slope being given by the spectral noise power at zero frequency.

Thus, the HEOM approach is not strictly needed in this context. However,
this compact analytic result provides the opportunity to compare the dynamics resulting from the HEOM approach with analytic results. In the case of pure decoherence, a double multi-index is not needed; the resulting hierarchy equation is
\begin{equation}\label{Eq:decoHEOM}
\begin{split}
\frac{d}{dt} \hat{\rho}_{\bf n} =
&%-\left(i\mathcal{L}_S
-\sum_{k=1}^{K}
n_{k} z_k %\right)
\hat{\rho}_{{\bf n}}\\ 
&-i\sum_{k=1}^{K}\sqrt{(n_k + 1)d_k} \left[\hat{q},\hat{\rho}_{{\bf n}_k^{+}} \right]  \\
&-i\sum_{k=1}^{K}\sqrt{n_k d_{k}}\;\left[\hat{q},\hat{\rho}_{{\bf n}_k^{-}} \right]
.
\end{split}
\end{equation}

Initially, we consider the simple case of a real correlation function with exponential time dependence,
%\begin{equation}
 $   C(t)= d e^{-\Gamma t}$.
%\end{equation}
This can be realized as a limiting case of an ohmic quantum reservoir with very weak coupling and very high temperature, or by replacing the reservoir with classical noise (Ornstein-Ulhlenbeck process). Within this section, we will assume $\Gamma>0$. Real $C(\tau)$ and positive noise power require $d$ to be positive as well, but we will relax this constraint here, since terms with negative or complex $d$ typically appear for low-temperature quantum reservoirs, described by more than one exponential term.
The corresponding HEOM system
\begin{equation}\label{eq:HEOMdephasing}
\begin{split}
\frac{d}{dt} \hat{\rho}_n =
-n \Gamma\hat{\rho}_n 
&-i\sqrt{(n + 1)d} \left[\hat{q},\hat{\rho}_{n+1} \right]\\
&-i\sqrt{n d} \left[\hat{q},\hat{\rho}_{n-1} \right]
\end{split}
\end{equation}
now has a single index, and we set $\rho_{-1}\equiv 0$.
We consider the time evolution of off-diagonal matrix elements of the physical density matrix and the ADOs, setting
$y_n(t) = \langle q|\hat\rho_n(t)| q'\rangle$, which results in

\begin{equation}\label{eq:HEOMdeph_scalar}
\begin{split}
\frac{d}{dt} y_n =
-n \Gamma y_n 
-i\sqrt{(n + 1)D}\, y_{n+1} 
-i\sqrt{n D}\, y_{n-1}
\end{split}
\end{equation}
with $y_{-1}\equiv 0$ and $D=d(q-q')^2$.
For a factorizing initial condition, $y_n(t=0)=0$ for $n>0$, this system has the closed-form solution
\begin{align}\label{eq:yn}
    y_n(t) =\frac{(-i)^n}{\sqrt{D^n n!}}\dot{\Phi}(t)^n \exp(-\Phi(t)) y_0(0)
\end{align}
with
\begin{align}
    \Phi(t) &= \frac{D}{\Gamma^2}(\Gamma t - 1 + e^{-\Gamma t})
%    \\
%    \dot\Phi(t) &= \frac{D}{\Gamma}(1-e^{-\Gamma t})\\
%    \ddot\Phi(t) &= D e^{-\Gamma t} = -\Gamma \dot\Phi + D
.
\end{align}
All of the $y_n(t)$ behave as $\propto e^{-Dt/\Gamma}$ in the long-time limit. Their transient behavior, which becomes markedly visible for $|D|\gg\Gamma^2$, indicates the non-Markovian nature of dephasing through colored noise. In the opposite limit, the coupling between hierarchy levels is a small perturbation, and hierarchy elements with $n>1$ are virtually negligible.

The right-hand side of the linear dynamical equation (\ref{eq:HEOMdeph_scalar}) defines a generator $G$ equivalent to the Hamiltonian of an inverted quantum harmonic oscillator with a complex shift,
\begin{equation}
    G = -\Gamma a^\dagger a - i\sqrt{D}(a^\dagger + a).
\end{equation}
Without any truncation, the eigenvalues of this generator can be obtained in complete analogy to the quantum mechanical oscillator; they are
\begin{equation}\label{eq:infeigen}
    \lambda_n  =  - n \Gamma - D/\Gamma.
\end{equation}
The analogy to the quantum oscillator is closest for the case $D<0$, where $G$ becomes hermitian with respect to the natural scalar product of sequence spaces. More generally, eq. (\ref{eq:infeigen}) is valid for arbitrary complex parameters $\Gamma$ and $D$, although the corresponding eigenvectors will generally be non-orthogonal. In particular, this is the case when both $D$ and $\Gamma$ are positive.

\subsection{Truncated HEOM dynamics}
The procedure of first obtaining exact solutions $y_n(t)$, then discussing which ones can be neglected is \emph{not}  equivalent to the approach which is taken when applying HEOM as a numerical algorithm: There, eq. (\ref{eq:HEOMdephasing}) must be truncated, typically by setting $\rho_n=0$ for $n$ larger than some $n_{\rm max}$. For a theoretical analysis, the truncation could also be imagined as a rank-one modification of the generator $G$ which makes it block diagonal, one block with $n\in [0,n_{\rm max}]$ and one (semi-infinite) block with $n>n_{\rm max}$.

\begin{figure*}
    \centering
    \includegraphics[width=0.7\textwidth]{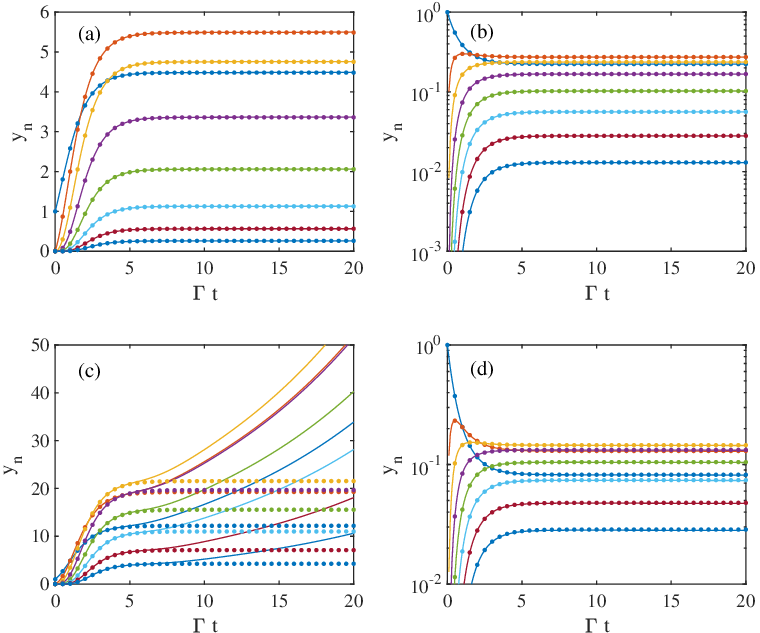}
    \vspace*{0.6eM}
    \caption{HEOM decoherence functions $y_n(t)\cdot e^{D t/\Gamma^2}$, normalized to $y_0(0)=1$ with 
    with and without truncation. Sign and strength of decoherence vary as a) $D=1.5\Gamma^2$, b) $D=-1.5\Gamma^2$, c) $D=2.5\Gamma^2$, d) $D=-2.5\Gamma^2$. The HEOM indices $0\ldots 7$ are color coded as dark blue, red, yellow, purple, green, light blue, cardinal and blue. Drawn lines represent HEOM result for finite truncation ($n_{\rm max}=10$); dots represent the exact result (\ref{eq:yn}).}
    \label{fig:preftime}
\end{figure*}

It is to be noted that there are cases in which the rank-one modification mentioned above is not a small perturbation: The level spacing of $G$ is $\Gamma$, and the modification eliminates matrix elements proportional to $\sqrt{n_{\rm max} D}$. The potential discrepancies between exact and truncated dynamics are illustrated in Fig. \ref{fig:preftime}, where the quantities $|y_n(t)| \exp(+Dt/\Gamma)$ are shown for $D = \pm 1.5 \Gamma^2$ and $D = \pm 2.5\Gamma^2$. Significant deviations are observed for large positive $D$, where decoherence effects are underestimated in HEOM. The different level of error depending on the sign of $D$ probably reflects the different nature of the eigensystem of $G$ for different signs of $D$ -- orthogonal vs. non-orthogonal.

\begin{figure*}
    \centering
    \includegraphics[width=0.6\textwidth]{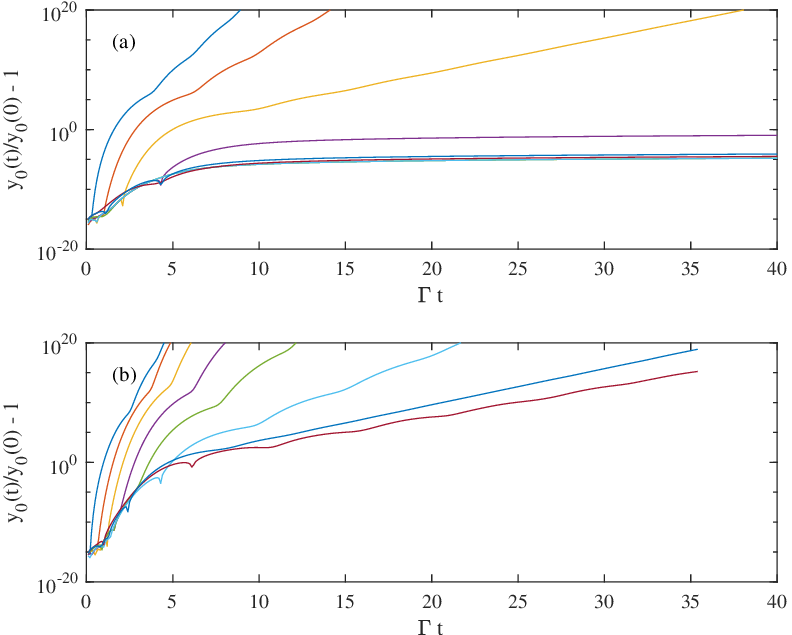}
    \vspace*{0.6eM}
    \caption{a) cancellation of opposing de- and recoherence terms for $D=10\Gamma^2$ and b) divergent error for $D=20\Gamma^2$.
    The maximum hierarchy index takes values 10, 20, 30, 40, 50, 60, 500 and 1000 (dark blue, red, yellow, purple, green, light blue, cardinal and blue).}
    \label{fig:nocanc}
\end{figure*}

When using a multi-exponential representation of a low-temperature quantum correlation function $C(t)$, one typically obtains one or more coefficients $d$ with negative real part, leading to \emph{rising} exponentials $e^{|D|t/\Gamma}$ in the resulting dynamics, which are compensated by other terms in the multi-exponential representation. Inaccuracies due to the hierarchy truncation can jeopardize this cancellation, as is occasionally observed in HEOM simulations~\cite{shi18,dunn19,yan20}. In order to reproduce this difficulty in a clean example, we consider the degenerate case of a formal two-exponential representation
\begin{equation}
C(t) = d e^{-\Gamma t} - d e^{-\Gamma t} = 0.
\end{equation}
The exact solution for $y_0(t)$ is obviously constant, $y_0(t) = y_0(0)$. Formally applying eq. (\ref{Eq:decoHEOM}) with $K=2$, $d_{1,2} = \pm d$ and $z_k=\Gamma$ results in a HEOM approximation of the exact result, the quality of which depends on the truncation parameter $n_{\rm max}$. The data in Fig. \ref{fig:nocanc} shows that a high truncation parameter can ensure this with a small realtive error of approximately $8\cdot 10^{-5}$ for a relatively large parameter $D = 10 \Gamma^2$ (a), while further raising $D$ to the value $20\Gamma^2$ (b) breaks the cancellation, leading to an exponential divergence even for extremely deep hierarchies.

\section{Stability issues in full HEOM dynamics}

\subsection{Time-domain analysis}
In practical HEOM computations, two slightly different truncation schemes are used. In the so-called N truncation scheme~\cite{li22}, all ADOs where \emph{any} index exceeds a given $n_{\rm max}$ are set to zero. In the L truncation scheme~\cite{li22}, ADOs where the \emph{sum} of all indices exceeds a given $n_{\rm max}$ are neglected. For the dynamics governed by eq. (\ref{Eq:baryheom4}), this yields $n_\textrm{max}^{2K}$ ADOs for N truncation and $\sum_{r = 0}^{n_\textrm{max}-1}\binom{2K-1+r}{r}$ ADOs for L truncation, a significantly smaller number than the former.
Converging the dynamics in a numerical simulation is simply a matter of increasing the truncation parameter $n_{\rm max}$ in either of the truncation methods until the dynamics no longer changes.
%Any simulation with a truncation parameter lower than the one required to reach convergence can be seen as as lower order pertubative dynamics. Thus, if we take the hierarchical cutoff $n_max$ to be analogous to a pertubative simulation of order $T$, we would expect the necessary hierarchy cutoff for convergent dynamics to have some linear relationship with the coupling strength $\alpha$ between system and reservoir.\\ 
In the following we investigate how this procedure succeeds or fails in the context of a spin-boson system with strong coupling and a sluggish, roughly resonant reservoir
with Hamiltonian $\hat{H}_s = \Delta \hat{\sigma}_x$ and coupling operator $\hat{q} = \hat{\sigma}_z$. For the influence of the reservoir a ohmic spectral density $J(\omega) = \alpha \omega /(1 + (\omega/\omega_c)^2)$ with Drude form cutoff was used. For a given coupling parameter $\alpha$, inverse temperature $\beta$ and cutoff frequency $\omega_c$ a barycentric fit~\cite{xu22,nakat20} with tolerance $10^{-2}$ was performed, yielding a correlation function with a single $(K = 1)$ complex exponential mode $C(t) = de^{-z t}$, which is more than adequate for a high-temperature reservoir.  The exploration of large $n_{\rm max}$ is thus facilitated.
\begin{figure*}
    \centering
    \includegraphics[width = 0.7\textwidth]{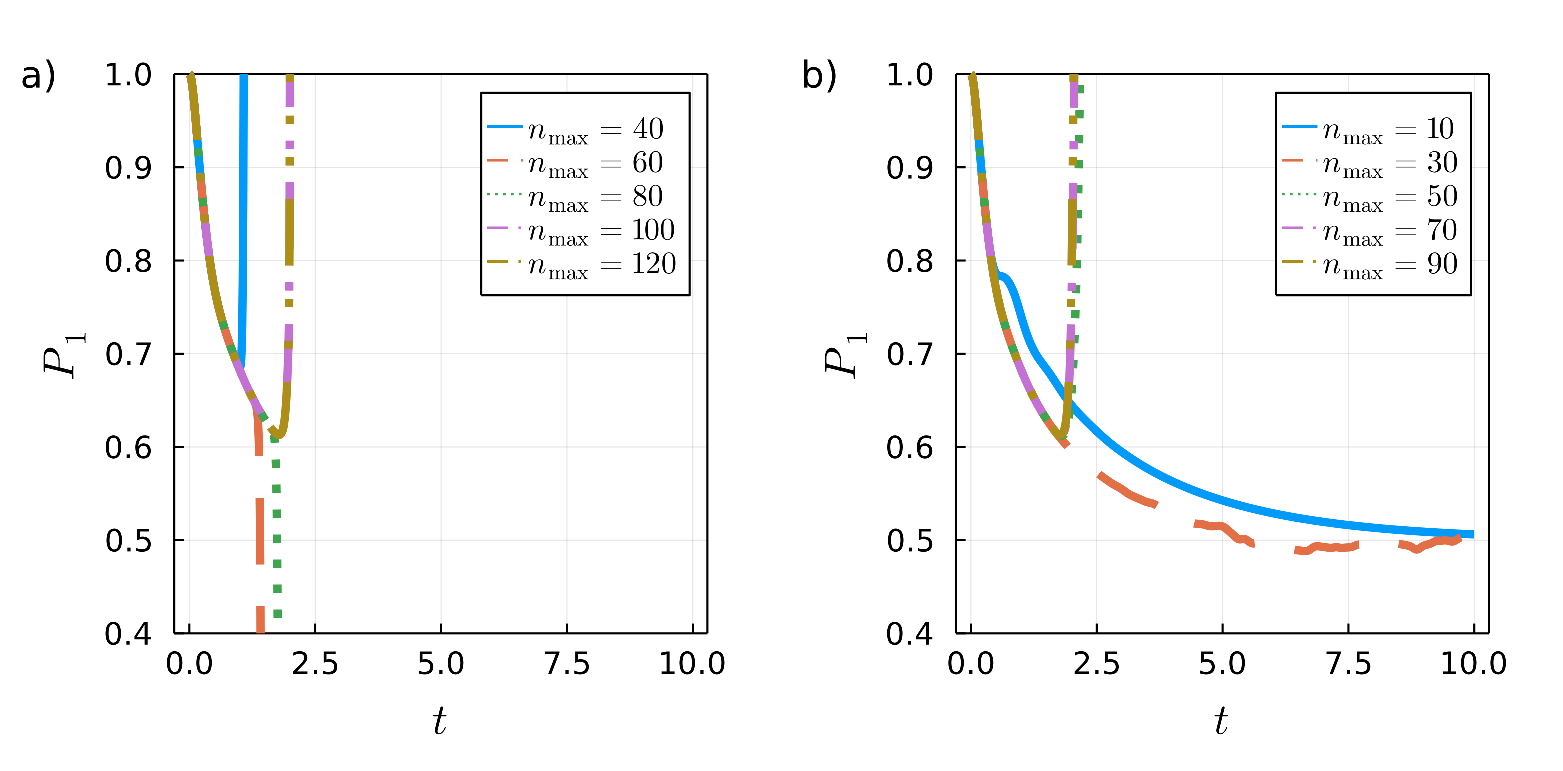}
    \caption{Heom example in the N-truncation a) and the L-truncation b) with problematic convergence properties for physical parameters $\omega_c = 1\,\mathrm{s}^{-1}$, $\alpha = 5$, $\omega_c \beta = 0.1$ and $\Delta = 4\,\mathrm{s}^{-1}$ for the spin-boson system with Hamiltonian $\hat{H}_s = \Delta\hat{\sigma}_x/2$ and coupling operator $\hat{q} = \hat{\sigma}_z$. The Barycentric-Fit, for an ohmic spectral density $J(\omega) = \alpha \omega/(1 + (\omega/\omega_c)^2)$ with Drude form cutoff, results in the HEOM parameters $d = 50.0 - 2.5\, i$ and $z = 1.0$}
    \label{fig:nconvtime}
\end{figure*}
In Fig.\ref{fig:nconvtime} the dynamics of the population of the exited state can be observed for the strong dissipation case $\alpha = 5$. The truncation parameter $n_\textrm{max}$ is steadily increased in an attempt to achieve physical dynamics for both the N- and L-Truncation. Within the numerically accessible parameter range, there appears to be pointwise convergence in the N-truncation up to a threshold near $n_\textrm{max} = 100$, with convergence attainable only on a finite time interval.

The population dynamics in the L-truncation displays somewhat different convergence properties. Low truncation orders do not seem to be unstable, but due to the strong dissipation also not yet converged. An increase of the truncation parameter eventually causes divergences instead of reaching a physical result; these divergences appear to persist for arbitrarily deep hierarchies.

\subsection{Spectral features of extended-state-propagation}
Since eq. (\ref{Eq:baryheom4}) is linear, HEOM instabilities can also be analyzed through the eigenvalues of the generator of the extended-state propagation.
For this purpose, it is convenient to vectorize the set of ADOs $\hat{\rho}_{{\bf m,n}}\rightarrow |\rho_{{\bf m,n}} \rangle$ and to define a global state $|\rho \rangle = (\dots, |\rho_{{\bf m,n}}\rangle, \dots)^t_{{\bf m,n}\in\mathcal{T}}$, where $\mathcal{T}$ is the set of all multi-indices $({\bf m,n})$ allowed by the truncation method. The FP-HEOM is now equivalently expressed by a linear map characterized by a Matrix $M(t)$, $d|\rho\rangle/dt = M(t)|\rho\rangle$. $M(t)$ is sparse, since it changes ADO indices at most by one. When it comes to investigating the long time stability, we will only consider the case of a time independent Hamiltonian, which allows the map to be expressed through a time-independent matrix $M$. From
\begin{equation}
    |\rho(t)\rangle = e^{M t} |\rho(0)\rangle
\end{equation}
it is obvious that an exponential divergence occurs when $M$ has an eigenvalue with positive real part.
Without simulating the system in the time domain we can simply inspect for such eigenvalues.

Determining the eigenvalues can be done either analytically or numerically~\cite{dunn19}. We use the ARPACK library~\cite{lehou98} for its efficient iterative algorithm in the search of eigenvalues with the largest real part for sparse matrices.
\begin{figure*}
    \centering
    \includegraphics[width=0.7\textwidth]{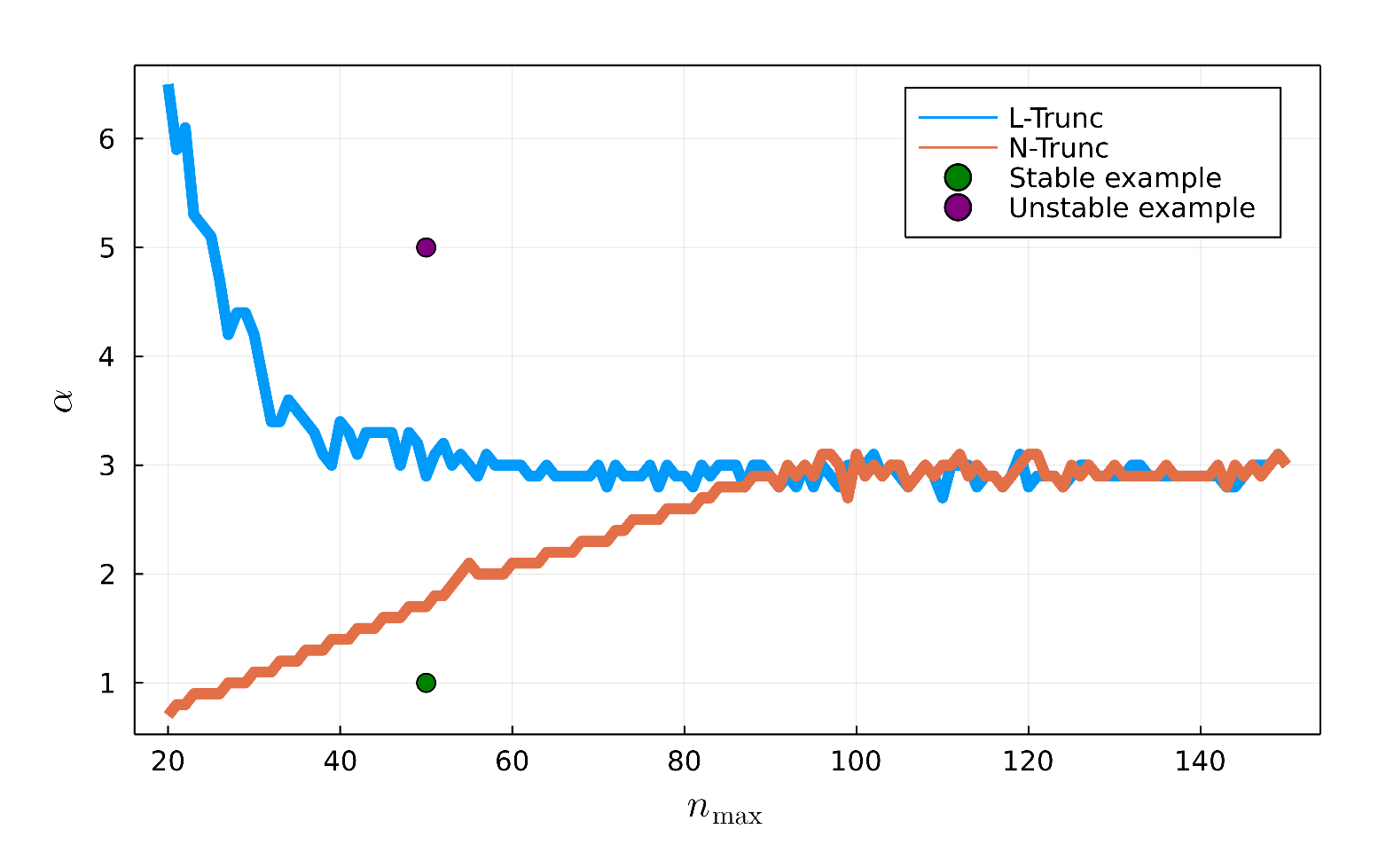}
    \caption{Boundary of convergent and divergent regime of the FP-HEOM. Physical parameters of the spin-boson system $\hat{H}_s = \Delta\hat{\sigma}_x/2$, $\hat{q} = \hat{\sigma}_z$ were chosen to be $\omega_c = 1\,\mathrm{s}^{-1}$, $\omega_c \beta = 0.1$ and $\Delta = 2\,\mathrm{s}^{-1}$. The barycentric fit was performed for all coupling parameters $\alpha$.}
    \label{fig:spectrum_comp}
\end{figure*}
For the numerical analysis of the instability we consider the same spin-boson system from the time-domain analysis. With stability thus defined, Fig.\ref{fig:spectrum_comp} shows a phase diagram with stable and unstable regions in the parameter plane spanned by $n_{\rm max}$ and $\alpha$, with different phase boundaries for L truncation and N truncation. In both cases, instabilities arise for $\alpha$ above the boundary. All $\alpha$ below the line will result in stable, but not necessarily converged dynamics.
In the L-truncation the allowed maximum coupling initially decreases with the depth of the hierarchy $n_\textrm{max}$ and tends towards some constant. For the N-truncation it seems to be the other way around, where the maximum allowed damping increases linearly with $n_\textrm{max}$ until some threshold depth and becoming constant afterwards.
From Fig.\ref{fig:spectrum_comp} it appears that for high hierarchy depths there is a critical coupling constant $\alpha_c \approx 3$ which becomes independent of $n_{\rm max}$. Thus, merely increasing the hierarchy depth $n_{\rm max}$ does not appear to be a viable strategy for curing divergence problems.

In the case of L truncation, comparably high values of $\alpha$ are allowed for smaller $n_{\rm max}$, which however, may be too small for numerical convergence of the dynamics. So in the worst case scenario, it might be impossible to converge the result without running into instabilities. This behaviour is consistent with the dynamics in Fig.\ref{fig:nconvtime}.

\begin{figure*}
    \centering
    \includegraphics[width = 0.7\textwidth]{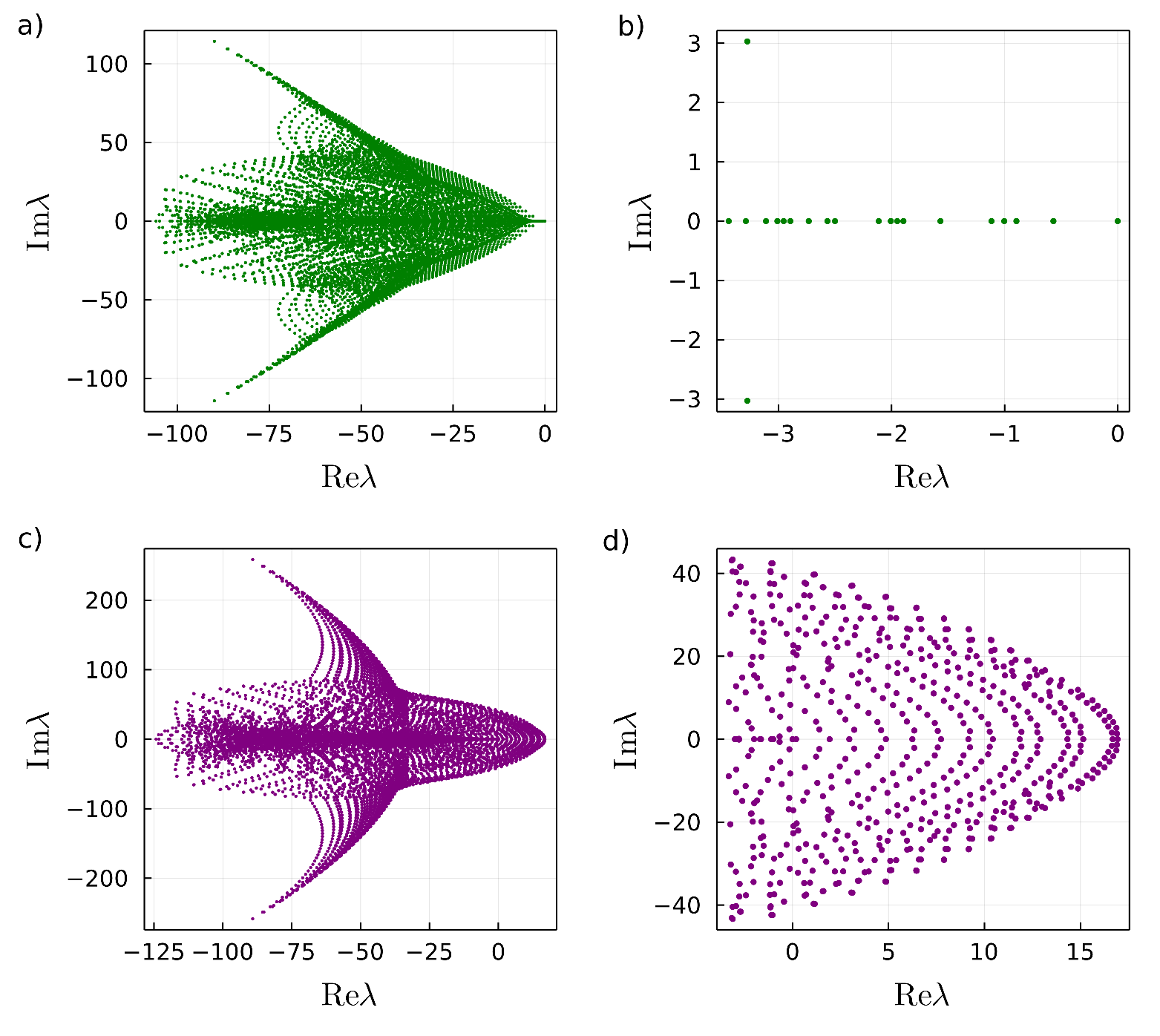}
    \caption{Full spectrum (left) and magnification (right) of the eigenvalues of the linear FP-HEOM map. Figures a) and b) correspond to the stable example points in Fig.\ref{fig:spectrum_comp}, and c) and d) correspond to the unstable example points respectively.}
    \label{fig:spec_alpha}
\end{figure*}

Fig.\ref{fig:spec_alpha} shows the full eigenvalue spectrum of two specific points for $n_\textrm{max} = 50$ in the N-truncation, namely $\alpha = 1$ and $\alpha = 5$. In the moderate damping case $\alpha = 1$ both the N- and L-truncation are stable and converged, whereas the strong damping case $\alpha = 5$ does not allow for stable dynamics. The spectrum for the stable case $\alpha = 1$ shows the absence of eigenvalues with positive real part and a eigenvalue with $\mathrm{Re}(\lambda) = 0$ corresponding to the steady state. In contrast, the spectrum for $\alpha = 5$ shows many eigenvalues with positive real part much greater than zero corresponding to unstable dynamics. For increasing $n_{\rm max}$, table \ref{tab:eig} shows the largest real part of any eigenvalue of $M$. It appears that these tend to a constant, again suggesting that a convergent and correct result cannot be achieved.

\begin{table*}
    \centering
    \begin{tabular}{|c|r|r|r|r|r|r|r|r|r|}
    \hline
         $\max {\rm Re}\lambda$ & 16.94 & 13.55 & 9.23 &
         5.21 & 5.12 & 5.25 & 5.10 & 5.22\\
    \hline
         $n_{\rm max}$ & 50 & 75 & 100 & 125 & 150 & 200 & 300 & 500\\
    \hline
    \end{tabular}
    \caption{Most unstable eigenvalue as a function of truncation depth. See text for details.}
    \label{tab:eig}
\end{table*}

Different remedies have been suggested against the problem of divergent cases.
Dunn et al.~\cite{dunn19} proposed a filtering algorithm which eliminates from $M$ the eigenvectors corresponding to unstable eigenvalues, with some success in the case where the modification of $M$ is of low rank. Considering the large number of eigenvalues with positive real part shown in Fig. \ref{fig:spec_alpha}d), with most real parts large compared to system parameters, the applicability of this procedure seems limited. Further development of truncation strategies approximating the omitted ADOs by a function of the retained ADOs~\cite{tanim91,ishiz05} instead of setting them to zero might lead to progress in solving this problem, possibly even allowing the use of smaller $n_{\rm max}$ than with a ``hard'' truncation. The reformulation of HEOM dynamics through a bath coordinate mapping~\cite{ikeda22} also seems a promising strategy in the case of a reservoir which is sluggish or near-resonant to the system. Finally, the combination of HEOM for fast reservoir components and an optimized stochastic approach for slow reservoir components~\cite{schmi19} can provide a solution to the problem.

\section{Conclusions}
The appearance of divergent errors in some HEOM computations is succinctly related to reservoir parameters and truncation depth. The simple, analytically solvable case of pure dephasing reveals that a hard truncation at some maximum index is not necessarily a perturbative change of the dynamics, and it does not become perturbative when the maximum index is raised. A numerical study of the spin-boson system clearly confirms this picture. Problems of this nature seem mostly confined to the combined regime of strong coupling and slow reservoir dynamics. Apart from this regime, the HEOM method remains a valid and efficient method, as evidenced by the vast majority of applications where divergences are not observed. For the problematic parameter regime, our work should stimulate the development of alternative strategies for the finite closure of HEOM equations or a HEOM mode structure less susceptible to divergence problems.

\section*{Acknowledgements}
We thank Meng Xu and Joachim Ankerhold for insightful and stimulating discussions. This work was supported through the SiQuRe project under the QT.BW program of the State of Baden-W\"urttemberg as well as the QSOLID project of BMBF (German Federal Ministry of Education and Research).

\section*{Data availability}
The data that support the figures within this article are available from the corresponding author upon reasonable request.

\bibliography{FQMT22,FQMT22m}% common bib file

\providecommand{\noopsort}[1]{}
\begin{thebibliography}{10}
\providecommand{\url}[1]{{#1}}
\providecommand{\urlprefix}{URL }
\providecommand{\doi}[1]{\url{https://doi.org/#1}}
\bibcommenthead

\bibitem{alick87}
R.~Alicki, K.~Lendi, \emph{Quantum Dynamical Semigroups and Applications},
  \emph{Lecture Notes in Physics}, vol. 286 (Springer, Berlin, 1987)

\bibitem{breue02}
H.P. Breuer, F.~Petruccione, \emph{The theory of open quantum systems} (Oxford
  University Press, Oxford, 2002), p. 625

\bibitem{weiss08b}
U.~Weiss, \emph{Quantum dissipative systems}, 3rd edn.
\newblock No.~13 in Series in modern condensed matter physics (World
  Scientific, Singapore, 2008), p. 507

\bibitem{diosi98}
L.~Di\'osi, N.~Gisin, W.T. Strunz, Non-{M}arkovian quantum state diffusion.
\newblock Phys. Rev. A \textbf{58}, 1699 (1998).
\newblock \doi{10.1103/PhysRevA.58.1699}

\bibitem{strun99b}
W.T. Strunz, L.~Di\'osi, N.~Gisin, T.~Yu, Quantum trajectories for {B}rownian
  motion.
\newblock Phys. Rev. Lett. \textbf{83}, 4909--4913 (1999)

\bibitem{stock01}
J.T. {Stockburger}, H.~{Grabert}, Non-{Markovian} quantum state diffusion.
\newblock Chemical Physics \textbf{268}, 249--256 (2001).
\newblock \doi{10.1016/S0301-0104(01)00307-X}

\bibitem{stock02}
J.T. Stockburger, H.~Grabert, Exact $c$-number representation of
  non-{M}arkovian quantum dissipation.
\newblock Phys. Rev. Lett. \textbf{88}, 170407 (2002).
\newblock \doi{10.1103/PhysRevLett.88.170407}

\bibitem{stock16a}
J.T. Stockburger, Exact propagation of open quantum systems in a
  system-reservoir context.
\newblock EPL (Europhysics Letters) \textbf{115}(4), 40010 (2016).
\newblock \doi{10.1209/0295-5075/115/40010}

\bibitem{tanim89}
Y.~Tanimura, R.~Kubo, Time evolution of a quantum system in contact with a
  nearly {Gaussian}-{Markoffian} noise bath.
\newblock J. Phys. Soc. Jpn. \textbf{58}(1), 101--114 (1989).
\newblock \doi{10.1143/JPSJ.58.101}

\bibitem{tanim91}
Y.~Tanimura, P.G. Wolynes, Quantum and classical {Fokker-Planck} equations for
  a {Gaussian}-{Markovian} noise bath.
\newblock Phys. Rev. A \textbf{43}(8), 4131--4142 (1991).
\newblock \doi{10.1103/PhysRevA.43.4131}

\bibitem{tanim06}
Y.~Tanimura, Stochastic {Liouville}, {Langevin}, {Fokker-Planck}, and master
  equation approaches to quantum dissipative systems.
\newblock J. Phys. Soc. Jpn. \textbf{75}, 082001 (2006).
\newblock \doi{10.1143/JPSJ.75.082001}

\bibitem{tanim20}
Y.~Tanimura, Numerically "exact" approach to open quantum dynamics: The
  hierarchical equations of motion ({HEOM}).
\newblock The Journal of Chemical Physics \textbf{153}(2), 020901 (2020).
\newblock \doi{10.1063/5.0011599}

\bibitem{xu22}
M.~Xu, Y.~Yan, Q.~Shi, J.~Ankerhold, J.T. Stockburger, Taming quantum noise for
  efficient low temperature simulations of open quantum systems.
\newblock Phys. Rev. Lett. \textbf{129}, 230601 (2022).
\newblock \doi{10.1103/PhysRevLett.129.230601}

\bibitem{ishiz06}
A.~Ishizaki, Y.~Tanimura, Modeling vibrational dephasing and energy relaxation
  of intramolecular anharmonic modes for multidimensional infrared
  spectroscopies.
\newblock J. Chem. Phys. \textbf{125}(8), 084501 (2006).
\newblock \doi{10.1063/1.2244558}

\bibitem{shi18}
Q.~Shi, Y.~Xu, Y.~Yan, M.~Xu, Efficient propagation of the hierarchical
  equations of motion using the matrix product state method.
\newblock The Journal of Chemical Physics \textbf{148}(17) (2018).
\newblock \doi{10.1063/1.5026753}

\bibitem{dunn19}
I.S. Dunn, R.~Tempelaar, D.R. Reichman, Removing instabilities in the
  hierarchical equations of motion: Exact and approximate projection
  approaches.
\newblock The Journal of Chemical Physics \textbf{150}(18) (2019).
\newblock \doi{10.1063/1.5092616}

\bibitem{yan20}
Y.~Yan, T.~Xing, Q.~Shi, A new method to improve the numerical stability of the
  hierarchical equations of motion for discrete harmonic oscillator modes.
\newblock The Journal of Chemical Physics \textbf{153}(20) (2020).
\newblock \doi{10.1063/5.0027962}

\bibitem{mengpriv}
\mbox{Meng} Xu, {private communication.}

\bibitem{li22}
T.~Li, Y.~Yan, Q.~Shi, A low-temperature quantum fokker-planck equation that
  improves the numerical stability of the hierarchical equations of motion for
  the brownian oscillator spectral density.
\newblock The Journal of Chemical Physics \textbf{156}(6), 064107 (2022).
\newblock \doi{10.1063/5.0082108}

\bibitem{nakat20}
Y.~Nakatsukasa, L.N. Trefethen, An algorithm for real and complex rational
  minimax approximation.
\newblock {SIAM} Journal on Scientific Computing \textbf{42}(5), A3157--A3179
  (2020).
\newblock \doi{10.1137/19M1281897}

\bibitem{lehou98}
R.B. Lehoucq, D.C. Sorensen, C.~Yang, \emph{ARPACK Users' Guide} (Society for
  Industrial and Applied Mathematics, Philadelphia, Pa., 1998).
\newblock \doi{10.1137/1.9780898719628}

\bibitem{ishiz05}
A.~Ishizaki, Y.~Tanimura, Quantum dynamics of system strongly coupled to
  low-temperature colored noise bath: Reduced hierarchy equations approach.
\newblock Journal of the Physical Society of Japan \textbf{74}(12), 3131--3134
  (2005).
\newblock \doi{10.1143/JPSJ.74.3131}

\bibitem{ikeda22}
T.~Ikeda, A.~Nakayama, Collective bath coordinate mapping of "hierarchy" in
  hierarchical equations of motion.
\newblock The Journal of Chemical Physics \textbf{156}(10), 104104 (2022).
\newblock \doi{10.1063/5.0082936}

\bibitem{schmi19}
K.~{Schmitz}, J.T. {Stockburger}, A variance reduction technique for the
  stochastic {Liouville-von Neumann} equation.
\newblock European Physical Journal Special Topics \textbf{227} (2019).
\newblock \doi{10.1140/epjst/e2018-800094-y}

\end{thebibliography}

\end{document}